\documentclass[conference]{IEEEtran}
\IEEEoverridecommandlockouts
\usepackage{cite}
\usepackage{amsmath,amssymb,amsfonts}
\usepackage{algorithmic}
\usepackage{graphicx}
\usepackage{textcomp}
\usepackage{comment}
\usepackage{amsthm}
\usepackage{calc}
\usepackage{stfloats}
\usepackage[printonlyused,nolist]{acronym}
\usepackage{xcolor}
\usepackage{setspace}
\def\BibTeX{{\rm B\kern-.05em{\sc i\kern-.025em b}\kern-.08em
    T\kern-.1667em\lower.7ex\hbox{E}\kern-.125emX}}

\theoremstyle{definition}
\newtheorem{definition}{Definition}

\begin{acronym}[acronym_list]
	\acro{BS}{base station}
	\acro{LoS}{line-of-sight}
	\acro{MIMO}{multiple-input multiple-output}
	\acro{OFDM}{orthogonal frequency division multiplexing}
	\acro{UE}{user equipment}
	\acro{NLoS}{non-line-of-sight}
	\acro{AoA}{angle-of-arrival}
	\acro{ToA}{time-of-arrival}
	\acro{PDP}{power delay profile}	
	\acro{GPS}{global positioning system}
	\acro{ULA}{uniform linear array}
	\acro{ADCM}{angle-delay channel matrix}
	\acro{DFT}{discrete Fourier transform}
	\acro{GRL}{gradient reversal layer}
	\acro{SGD}{stochastic gradient descent}
	\acro{SA}{self-attention}
	\acro{QuaDRiGa}{Quasi Deterministic Radio Channel Generator}
	\acro{CNN}{convolutional neural network}
	\acro{DNN}{deep neural network}
	\acro{FBS}{first-bounce scaterrer}
	\acro{LBS}{last-bounce scaterrer}
	\acro{CDF}{cumulative distribution function}
\end{acronym}    

\newlength{\depthofsumsign}
\newlength{\totalheightofsumsign}
\newlength{\heightanddepthofargument}

\newcommand{\nsum}[1][1.4]{
    \mathop{%
        \raisebox
            {-#1\depthofsumsign+1\depthofsumsign}
            {\scalebox
                {#1}
                {$\displaystyle\sum$}%
            }
    }
}

\begin{document}
\title{Localization in Dynamic Indoor MIMO-OFDM Wireless Systems using Domain Adaptation
\thanks{This work is partially supported by the Federal Ministry of Education and Research of Germany in the programme of “Souver{\"a}n. Digital. Vernetzt.” Joint project 6G-RIC (Project ID: 16KISK030), and 6G-ANNA project (Project ID: 16KISK101).}}

\author{\IEEEauthorblockN{Rafail Ismayilov}
	\IEEEauthorblockA{\textit{Fraunhofer Heinrich Hertz Institute} \\
		Berlin, Germany \\
	rafail.ismayilov@hhi.fraunhofer.de} \vspace{-21pt}
	\and
	\IEEEauthorblockN{Renato L. G. Cavalcante}
	\IEEEauthorblockA{\textit{Fraunhofer Heinrich Hertz Institute} \\
		\textit{Technical University of Berlin}\\
		Berlin, Germany \\
	renato.cavalcante@hhi.fraunhofer.de} \vspace{-21pt}
	\and
	\IEEEauthorblockN{Sławomir Stańczak}
	\IEEEauthorblockA{\textit{Fraunhofer Heinrich Hertz Institute} \\
		\textit{Technical University of Berlin}\\
		Berlin, Germany \\
	slawomir.stanczak@hhi.fraunhofer.de} \vspace{-21pt}
}
\maketitle

\makeatletter
\renewcommand\footnoterule{%
	\kern-3\p@
	\hrule\@width.4\columnwidth
	\kern2.6\p@}
	
\newcommand*{\DivideLengths}[2]{%
	\strip@pt\dimexpr\number\numexpr\number\dimexpr#1\relax*65536/\number\dimexpr#2\relax\relax sp\relax
}

\makeatother
  
\begin{abstract}
	We propose a method for predicting the location of \ac{UE} using wireless fingerprints in dynamic indoor \ac{NLoS} environments.
	In particular, our method copes with the challenges posed by the drift, birth, and death of scattering clusters resulting from dynamic changes in the wireless environment.
	Prominent examples of such dynamic wireless environments include factory floors or offices, where the geometry of the environment undergoes changes over time.
	These changes affect the distribution of wireless fingerprints, demonstrating some similarity between the distributions before and after the change.
	Consequently, the performance of a location estimator initially designed for a specific environment may degrade significantly when applied after changes have occurred in that environment.
	To address this limitation, we propose a domain adaptation framework that utilizes neural networks to align the distributions of wireless fingerprints collected both before and after environmental changes.
	By aligning these distributions, we design an estimator capable of predicting \ac{UE} locations from their wireless fingerprints in the new environment.
	Experiments validate the effectiveness of the proposed methods in localizing \acp{UE} in dynamic wireless environments.
\end{abstract}

\vspace{-8pt}
\section{Introduction}
\label{sec:intro}
\vspace{-5pt}
\acresetall 
The integration of wireless fingerprint localization into future wireless communication systems is gaining increasing interest due to its potential applications.
Promising applications of wireless fingerprint localization are indoors, where global positioning system provides poor localization performance or fails to localize \ac{UE} due to the absence of line-of-sight propagation from satellites to the \ac{UE}.
Another promising application of wireless fingerprint localization is beamforming enhancement in millimeter-wave wireless communication systems.
For example, the proposed method in \cite{Ismayilov_2021_2} relies entirely on \ac{UE} location information to predict the beams, while the proposed method in \cite{Ismayilov_2021_1} exploits the \ac{UE} location as side information to improve the beamforming performance.
While location information can enhance the performance of wireless systems, obtaining it from channel measurements, in general, is challenging.
Existing methods in the literature \cite{Garcia_2017, Sun_2019, Song_2022} use neural networks for wireless fingerprint localization.
These methods exploit the fact that the wireless channel between the \ac{UE} and \ac{BS} is strongly related to the scattering environments for a given \ac{UE} location.
This property allows us to use neural networks to extract features from channel measurements that are strongly related to each \ac{UE} location, further posing the localization problem as a pattern recognition problem.
\begin{figure}[!t]
	\centering
	\centerline{\includegraphics[width=84.0mm]{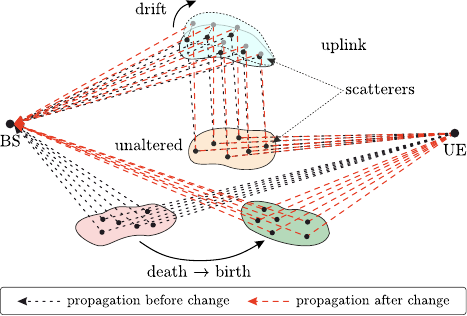}}
	\vspace{-7pt}
	\caption{Illustration of a dynamic wireless environment where some scattering clusters drift, some remain unaltered, and others die, while new ones are born due to environmental changes. Consequently, some parameters (e.g., \ac{AoA} and latency) in wireless fingerprints before and after the environmental change exhibit similarities.}
	\label{fig.environment_change_illustration}
	\vspace{-20pt}
\end{figure}

In practical wireless systems, the environment is often non-stationary owing to changes related to object movement or rearrangement. 
These changes typically result in the birth, death and drift of scattering clusters, as illustrated Fig.~\ref{fig.environment_change_illustration}.
Consequently, previously learned features become obsolete for localization in the environment after these changes, so we need to cope with changes in the distribution of wireless channels for localization in dynamic environments.
In the case of drastic changes in the distribution, the problem is very challenging. \footnote{With drastic changes in the distribution of wireless fingerprints, the location estimator can be retrained with new samples, or conventional localization methods can be applied.}
However, according to experimental measurements \cite{3GPP_2018_06}, some parameters of wireless channels corresponding to the same \ac{UE} location before and after environmental changes exhibit similarities in practical wireless systems.
These similarities manifest in parameters such as \ac{AoA}, power delay profile, or a combination of both, which are of particular interest for wireless fingerprint localization, as the \ac{MIMO}-\ac{OFDM} system enables the \ac{BS} to resolve the \ac{AoA} by exploiting the antenna arrays and leverage the delay information extracted from the \ac{OFDM} signals.

Against this background, we propose methods that exploit the similarity of wireless fingerprints to localize \acp{UE} in a dynamic environment.
Specifically, we consider environments before and after changes as two distinct domains, referred to as the source domain and the target domain, respectively. 
Employing a domain adaptation approach, we aim to learn the similarity between these domains. 
To this end, we introduce two neural network architectures.
The first architecture is based on autoencoders, where features extracted by the encoder from wireless fingerprints in both domains are expected to follow similar distributions.
Consequently, a location estimator trained in one domain can effectively localize \acp{UE} in another domain. 
The second architecture employs adversarial learning to capture features discriminative for the localization task in the source domain and invariant to distribution changes between the source and target domains. 
In both architectures, domain adaptation is performed in an unsupervised manner, eliminating the need to access \ac{UE} location information after environmental changes, which is critical in practical wireless communication systems.
We evaluate the performance of the proposed architectures with simulations.
\section{System Model and Problem Statement}
\vspace{-1pt}
\subsection{System Model}
\label{system_model}
In the following, we consider an uplink wireless \ac{MIMO}-\ac{OFDM} communication system consisting of a single \ac{BS} and multiple \acp{UE}.
The \ac{BS} is assumed to use a \ac{ULA} with $L$ antenna elements, each spaced $\lambda / 2$ apart from each other, where $\lambda$ denotes the wavelength of the carrier frequency.
The \acp{UE} are equipped with a single antenna.
The wireless channel obtained at the \ac{BS} from an arbitrary \ac{UE} location is denoted by $\widetilde{\mathbf{H}} \in \mathbb{C}^{L \times K}$, where $K$ indicates the number of \ac{OFDM} subcarriers.
In this work, we use the obtained wireless channel $\widetilde{\mathbf{H}}$ to construct an \ac{ADCM}, which we subsequently refer to as a wireless fingerprint.
The \ac{ADCM} (denoted by $\mathbf{H}$) is derived from $\widetilde{\mathbf{H}}$ as follows \cite{Sun_2018}:
\vspace{-4pt}
\begin{equation}\label{eq.ADCM_fingerprint}
	\mathbf{H} = \mathbf{V}^H \widetilde{\mathbf{H}} \mathbf{F},
	\vspace{-5pt}
\end{equation}
where $\mathbf{V} \in \mathbb{C}^{L \times L}$ is a phase shifted \ac{DFT} matrix with its components $\left [ \mathbf{V} \right ]_{i,j}$ of the $i^{\text{th}}$ row and $j^{\text{th}}$ column given by $\left [ \mathbf{V} \right ]_{i,j} \triangleq \frac{1}{\sqrt{L}} e^{ -j 2 \pi \frac{i \left ( j - \frac{L}{2} \right )}{L} }$, and $\mathbf{F} \in \mathbb{C}^{K \times K}$ denotes the unitary \ac{DFT} matrix with elements given by $\left [\mathbf{F} \right ]_{i,j} \triangleq \frac{1}{\sqrt{K}} e^{ -j 2 \pi \frac{ij}{K} }$. 
Given the assumption of utilizing \ac{ULA}, the matrices $\mathbf{V}$ and $\mathbf{F}$ in \eqref{eq.ADCM_fingerprint} map the space-frequency domain to the angle and delay domains, respectively.
Thus, the $ \left ( i,j \right )^{\text{th}}$ element of $\mathbf{H}$ represents the complex gain associated with the $i^{\text{th}}$ \ac{AoA} and the $j^{\text{th}}$ time-of-arrival with respect to a given reference time point.
In a wireless environment with dynamic scatterers, we assume that the \ac{ADCM} obtained before and after the environment change are expected to maintain some similarity.
The study in \cite{3GPP_2018_06} provides us with empirical evidence that this assumption can be satisfied in some scenarios of practical relevance.
\begin{figure}[t!]
	\centering
	\centerline{\includegraphics[width=88.0mm]{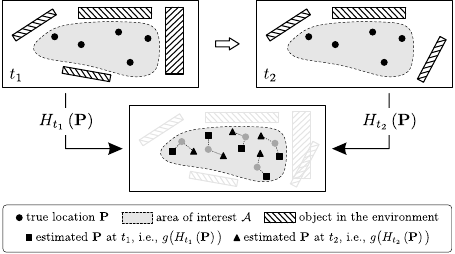}}	
	\caption{Illustration of similarity in dynamic environments (Definition~\ref{def.env_similarity}).}
	\label{fig.def1_illustration}	
	\vspace{-10pt}
\end{figure}
\subsection{Problem Statement}
Until this point, we have utilized the notion of similarity to abstractly express the relation between environments before and after changes.
In general, defining similarities between environments is not trivial, particularly in complex indoor \ac{NLoS} environments.
One possibility to define similarity between environments might involve considering the proportion of scattering clusters that remain unaltered after a change in the environment, e.g., only 10$\%$ of the scattering clusters experience alteration.
However, even a small number of scattering clusters may encapsulate highly significant localization features, and their birth, death, drift, or size in a dynamic environment can substantially affect localization performance.
Conversely, the birth, death, drift, or size of numerous other scattering clusters may have negligible impact on localization performance.
In this work, we propose the following definition of similarity between environments for a given estimator mapping \acp{ADCM} to locations.
\begin{definition}[similarity of environments]\label{def.env_similarity}
	Let $t_1 \in \mathbb{R}$ and $t_2 \in \mathbb{R}$ denote two time points such that $t_1 < t_2$.
	Let $\mathbf{P} \in \mathcal{A}$ denote the location of a \ac{UE} within the area of interest $\mathcal{A} \subseteq \mathbb{R}^2$.
	Let $H_{t}  :\mathbb{R}^2 \rightarrow \mathbb{C}^{L \times K}$ be the function that maps $\mathbf{P}$ to the wireless fingerprints at time $t$.
	Furthermore, let $g:\mathbb{C}^{L \times K} \rightarrow \mathbb{R}^2$ be a given estimator that maps fingerprints to locations, and assume that there exists $Q > 0$ such that $\left ( \forall t \in \mathbb{R} \right ) \left ( \forall \mathbf{P} \in \mathcal{A} \right ) \left\| g \left ( H_{ t } \left ( \mathbf{P} \right ) \right )\right\| \le Q$, where $\left\| \cdot \right\|$ represents the standard $l_2$ norm.
	Then, the similarity of the environments at time points $t_1$ and $t_2$ for the estimator $g$ is given by:
	\begin{equation}
		\sigma_g \left ( t_1, t_2 \right ):= \underset{\mathbf{P} \in \mathcal{A}}{\text{sup}} \left\| g \big ( H_{t_1 } \left ( \mathbf{P} \right ) \big ) - g \big ( H_{ t_2 } \left ( \mathbf{P} \right ) \big ) \right\|,
	\end{equation}
	where sup denotes the supremum operator.
\end{definition}

To gain an intuition of the similarity proposed in Definition~\ref{def.env_similarity}, we present a simple illustration in Fig~\ref{fig.def1_illustration}.
In Definition~\ref{def.env_similarity}, $\sigma_g \left ( t_1, t_2 \right ) \geq 0$, and ideally, with stronger similarity of the environments at time points $t_1$ and $t_2$, $\sigma_g \left ( t_1, t_2 \right )$ should approach zero, while it should increase with decreasing similarity of the environments at time points $t_1$ and $t_2$.
However, note that although this property is desirable, it is not guaranteed for every $g$.
Moreover, in practice, we need numerical approximations to compute the similarity of the environments for time points for a given estimator $g$ in the sense of Definition~\ref{def.env_similarity}.
Next, we formulate the wireless fingerprint localization problem in a dynamic environment as a domain adaptation problem as follows.

Let $\mathcal{S} = \left ( \mathcal{H}, \mathcal{P}, \pi_{S} \right )$ be the source domain corresponding to the environment at time point $t_1$, and let $\mathcal{T} = \left ( \mathcal{H}, \mathcal{P}, \pi_{T} \right )$ be the target domain corresponding to the environment at time point $t_2$, where $\mathcal{H}$ represents the wireless fingerprint space, $\mathcal{P}$ represents the \ac{UE} location space, $\pi_{S}$ and $\pi_{T}$ denote the distributions on $\mathcal{H} \times \mathcal{P}$ for the source and target domains, respectively.
We assume that the distributions $\pi_{S}$ and $\pi_{T}$ are different. 
Thus, the source domain represents the wireless environment before the change, and the target domain represents the wireless environment after the change.

In practice, only a finite number of wireless fingerprints can be acquired in an environment.
This leads us to consider sets containing pairs of fingerprints and corresponding \ac{UE} locations collected in both the source and target domains.
We denote the set containing the fingerprint-location pairs collected in the source domain as $ D_S = \left\{ \left ( \mathbf{H}_i, \mathbf{P}_i \right ) \right\}_{i =1}^{N_S}$ with $\left ( \forall i \in  \left\{1,...,N_S \right\} \right ) \left ( \mathbf{H}_i, \mathbf{P}_i \right ) \sim \pi_S$, and we denote the set containing the fingerprint-location pairs collected in the target domain as $ D_T = \left\{ \left ( \mathbf{H} _i, \mathbf{P}_i \right ) \right\}_{i=1}^{N_T}$ with $\left ( \forall i \in  \left\{1,...,N_T \right\} \right ) \left ( \mathbf{H}_i, \mathbf{P}_i \right ) \sim \pi_T$, where $N_S$ and $N_T$ denote the number of samples collected in source and target domains, respectively.
However, in many cases, particularly after a change in the environment, we often have access only to wireless fingerprints without the corresponding \ac{UE} locations. 
That is, we exclude \ac{UE} locations from the set $D_T$, resulting in another set $\bar{D}_T = \left\{ \left ( \mathbf{H}_i\right ) \right\}_{i=1}^{N_T}$ containing only the wireless fingerprints in the target domain. 
For further use in the next section, we similarly construct the set $\bar{D}_S = \left\{ \left ( \mathbf{H}_i\right ) \right\}_{i=1}^{N_S}$ containing only the wireless fingerprints in the source domain.
Now, given $D_S$ and $\bar{D}_T$, the objective of domain adaptation is to exploit these sets to find an estimator $g$ that minimizes the localization error in the target domain.
Ideally, this requires us to solve the following problem:
\begin{equation}\label{eq.problem_statement}
	\underset{g \in \mathcal{G}}{\text{minimize}} \hspace{1pt} \underset{\left ( \mathbf{H}, \mathbf{P} \right ) \sim \pi_T}{\mathbb{E}} \Big [ \big \| g \left ( \mathbf{H}; D_S, \bar{D}_T\right ) -  \mathbf{P} \big \| \Big ] ,
\end{equation}
\begin{figure*}[!hb]
\vspace{-10pt}
	\hrule
	\vspace{5pt}
	\begin{equation}\label{eq.autoencoder_opt}
		\left ( \boldsymbol{\alpha}^{\star}, \boldsymbol{\beta}^{\star}, \boldsymbol{\gamma}^{\star} \right ) \in \underset{\boldsymbol{\alpha}, \boldsymbol{\beta}, \boldsymbol{\gamma}}{\text{arg min}} \Bigg [ \dfrac{1}{N_S}\displaystyle\nsum[1.3]_{i=1}^{N_S}  \Big\| \underbrace{ g_{\text{AE}}^{\boldsymbol{\gamma}} \big ( f_{\text{AE}}^{\boldsymbol{\alpha}} \left ( \mathbf{H}_i \right ) \big ) }_{\widehat{\mathbf{P}}_i} - \mathbf{P}_i \Big\| + \dfrac{1}{N_S + N_T}\displaystyle\nsum[1.3]_{i=1}^{N_S + N_T}  \Big\| \underbrace{\widehat{f}_{\text{AE}}^{\boldsymbol{\beta}} \big ( f_{\text{AE}}^{\boldsymbol{\alpha}} \left ( \mathbf{H}_i \right ) \big ) }_{\widehat{\mathbf{H}}_i} - \mathbf{H}_i \Big\|  \Bigg ] 
	\end{equation}
	\begin{equation}\label{eq.L_g_GR}
		\resizebox{0.95\hsize}{!}{$
			\mathcal{L}_{\text{GR}} \left ( \boldsymbol{\delta}, \boldsymbol{\epsilon}, \boldsymbol{\zeta} \right ) = \dfrac{1}{N_S} \displaystyle\nsum[1.5]_{i=1}^{N_S}\bigg \| \underbrace{ g_{\text{GR}}^{\boldsymbol{\zeta}} \left ( f_{\text{GR}}^{\boldsymbol{\delta}} \left ( \mathbf{H}_i \right ) \right ) }_{\widehat{\mathbf{P}}_i} - \mathbf{P}_i \bigg\|  + \dfrac{\lambda}{N_S + N_T} \displaystyle\nsum[1.5]_{i=1}^{N_S + N_T} \bigg[ d_i \cdot \log \Big ( \underbrace{h_{\text{GR}}^{\boldsymbol{\epsilon}} \big ( f_{\text{GR}}^{\boldsymbol{\delta}} \left ( \mathbf{H}_i \right )\big )}_{\widehat{d}_i} \Big ) + \left ( 1-d_i \right ) \cdot \log \Big (1- \underbrace{h_{\text{GR}}^{\boldsymbol{\epsilon}} \big ( f_{\text{GR}}^{\boldsymbol{\delta}} \left ( \mathbf{H}_i \right )\big )}_{\widehat{d}_i} \Big ) \bigg]
		$}
	\end{equation}
\end{figure*}
\hspace{-8.5pt} where $g$ is parameterized by $D_S$ and $\bar{D}_T$, $\mathbb{E}$ stands for the expectation, and $\mathcal{G}$ is a suitably chosen class of functions.
Solving the problem in \eqref{eq.problem_statement} involves challenges associated with computing the expectation over the unknown distribution $\pi_T$. 
To address these challenges, we employ regularized empirical mean minimization, in which the expectation over $\pi_T$ is replaced with an average computed from the observed data, and the regularizer imposes the extraction of domain-invariant features.
In particular, we consider two regularized empirical means provided in \eqref{eq.autoencoder_opt} and \eqref{eq.L_g_GR} at the bottom of this page, and in the following, we provide methods for minimizing them and describe their relation to \eqref{eq.problem_statement}.

\section{Localization with Unsupervised Domain Adaptation}
\begin{figure*}[htb]
	\centering
	\centerline{\includegraphics[width=180mm]{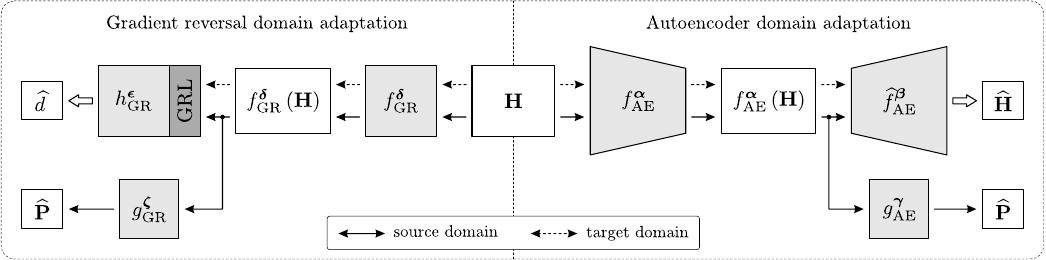}} 
	\caption{The proposed architectures for predicting the \ac{UE} locations from their wireless fingerprints in dynamic wireless \ac{NLoS} environments.}
	\label{fig.proposed_methods_diagram}
	\vspace{-10pt}
\end{figure*}
\subsection{Domain Adaptation with Autoencoders}\label{sec.DA_autoenc}
We begin this section by describing the relation between \eqref{eq.problem_statement} and \eqref{eq.autoencoder_opt}. 
Specifically, the components involved in \eqref{eq.autoencoder_opt} are: the encoder $f_{\text{AE}}^{\boldsymbol{\alpha}} : \mathbb{C}^{L \times K} \rightarrow \mathbb{R}^B$ parameterized by $\boldsymbol{\alpha}$ (where $B$ denotes the output dimension of the encoder), decoder $\widehat{f}_{\text{AE}}^{\boldsymbol{\beta}}: \mathbb{R}^B \rightarrow \mathbb{C}^{L \times K}$ parameterized by $\boldsymbol{\beta}$, and estimator $g_{\text{AE}}^{\boldsymbol{\gamma}}: \mathbb{R}^B \rightarrow \mathbb{R}^2$ parameterized by $\boldsymbol{\gamma}$.
The architecture containing the listed components above is illustrated in Fig.~\ref{fig.proposed_methods_diagram} (right).
The encoder $f_{\text{AE}}^{\boldsymbol{\alpha}}$ is designed to extract features from wireless fingerprints. 
The decoder $\widehat{f}_{\text{AE}}^{\boldsymbol{\beta}}$ has a symmetric structure with the encoder $f_{\text{AE}}^{\boldsymbol{\alpha}}$, and it is designed to output $\widehat{\mathbf{H}}$, a reconstruction of the inputs, from the features $f_{\text{AE}}^{\boldsymbol{\alpha}} \left (  \mathbf{H} \right )$, i.e., $\widehat{\mathbf{H}} = \widehat{f}_{\text{AE}}^{\boldsymbol{\beta}} \left ( f_{\text{AE}}^{\boldsymbol{\alpha}} \left ( \mathbf{H} \right )\right ) $.
The estimator $g_{\text{AE}}^{\boldsymbol{\gamma}}$ takes the output of the encoder $f_{\text{AE}}^{\boldsymbol{\alpha}} \left (  \mathbf{H} \right )$ as input, and it produces an estimated location $\widehat{\mathbf{P}}$ of the \acp{UE} as output, i.e., $\widehat{\mathbf{P}} = g_{\text{AE}}^{\boldsymbol{\gamma}} \left (  f_{\text{AE}}^{\boldsymbol{\alpha}} \left (  \mathbf{H} \right ) \right ) $. 
Now, the first term in \eqref{eq.autoencoder_opt} represents the localization error similar to the error in \eqref{eq.problem_statement}, and the second term acts as a regularizer. 
That is, $g_{\text{AE}}^{\boldsymbol{\gamma}} \circ f_{\text{AE}}^{\boldsymbol{\alpha}}$ in \eqref{eq.autoencoder_opt} provides location estimates similar to $g$ in \eqref{eq.problem_statement}. 
Furthermore, the relation between \eqref{eq.autoencoder_opt} and \eqref{eq.problem_statement} is that minimizing the empirical mean in \eqref{eq.autoencoder_opt} approximates $g$ in \eqref{eq.problem_statement}, and the regularizer in \eqref{eq.autoencoder_opt} enforces the extraction of common features from both domains in the provided architecture.

We learn the parameters $\boldsymbol{\alpha}^{\star}$, $\boldsymbol{\beta}^{\star}$, $\boldsymbol{\gamma}^{\star}$ (under the assumption of their existence) as specified in \eqref{eq.autoencoder_opt}.
In particular, the parameters $\boldsymbol{\alpha}^{\star}$ together with $\boldsymbol{\beta}^{\star}$ are learned using wireless fingerprint samples from both $\bar{D}_S$ and $\bar{D}_T$, i.e., $\left ( \forall i \in  \left\{1,...,N_S + N_T \right\} \right ) \mathbf{H}_i \in \bar{D}_S \cup \bar{D}_T$, by minimizing the $l_2$ distance between the true wireless fingerprints $\mathbf{H}_i$ and their reconstructions $\widehat{\mathbf{H}}_i = \widehat{ f}_{\text{AE }}^{\boldsymbol{\beta}} \left ( f_{\text{AE}}^{\boldsymbol{\alpha}} \left ( \mathbf{H}_i \right ) \right ) $.
Additionally, the parameters $\boldsymbol{\alpha}^{\star}$ together with $\boldsymbol{\gamma}^{\star}$ are learned using wireless fingerprint - locations pairs from $D_S$, i.e., $\left ( \forall i \in  \left\{1,...,N_S\right\} \right ) \left ( \mathbf{H}_i, \mathbf{P}_i \right ) \in D_S$, by minimizing the $l_2$ distance between true \ac{UE} locations in the source domain $\mathbf{P}_i$ and their estimates $\widehat{\mathbf{P}}_i = g_{\text{AE}}^{\boldsymbol{\gamma}} \left (  f_{\text{AE}}^{\boldsymbol{\alpha}} \left (  \mathbf{H}_i \right ) \right ) $.
In this way, the encoder seeks the parameter $\boldsymbol{\alpha}^{\star}$ that enables it to extract those features from wireless fingerprints that are common to both domains, while also allowing the estimator $g_{\text{AE}}^{\boldsymbol{\gamma}^{\star}}$ to utilize those features to localize \acp{UE} in the source domain.
That is, the extracted features from wireless fingerprints from both domains are expected to follow the same distribution.
This allows us to discard $\widehat{f}_{\text{AE}}^{\boldsymbol{\beta}^{\star}}$ and use $f_{\text{AE}}^{\boldsymbol{\alpha}^{\star}}$ together with $g_{\text{AE}}^{\boldsymbol{\gamma}^{\star}}$ to predict the \ac{UE} locations in the target domain from their wireless fingerprints.
That is, given the problem in \eqref{eq.problem_statement}, we obtain 
\begin{equation*}
	g \approx g_{\text{AE}}^{\boldsymbol{\gamma}^{\star}} \circ  f_{\text{AE}}^{\boldsymbol{\alpha}^{\star}}.
\end{equation*}
\subsection{Domain Adaptation with Gradient Reversing}\label{sec.DA_GRL}
We begin this section by describing the relation between \eqref{eq.problem_statement} and \eqref{eq.L_g_GR}. 
Specifically, the components involved in \eqref{eq.L_g_GR} are: the feature extractor $f_{\text{GR}}^{\boldsymbol{\delta}} : \mathbb{C}^{L \times K} \rightarrow \mathbb{R}^B$ parameterized by $\boldsymbol{\delta}$ (where $B$ denotes the output dimension of the feature extractor), domain classifier $h_{\text{GR}}^{\boldsymbol{\epsilon}} : \mathbb{R}^B \rightarrow  \left\{ 0,1\right\}$ parameterized by $\boldsymbol{\epsilon}$, and estimator $g_{\text{GR}}^{\boldsymbol{\zeta}}: \mathbb{R}^B \rightarrow \mathbb{R}^2$ parameterized by $\boldsymbol{\zeta}$.
The architecture containing the listed components above is illustrated in Fig.~\ref{fig.proposed_methods_diagram} (left).
The feature extractor $f_{\text{GR}}^{\boldsymbol{\delta}}$ is designed to extract features from wireless fingerprints. 
The domain classifier $h_{\text{GR}}^{\boldsymbol{\epsilon}}$ is designed to output $\widehat{d}$, a prediction of the domain labels $d \in \left\{ 0,1\right\}$ (i.e., source or target domain samples), from the extracted features $f_{\text{GR}}^{\boldsymbol{\delta}} \left (  \mathbf{H} \right )$, i.e., $\widehat{d} = h_{\text{GR}}^{\boldsymbol{\epsilon}} \left ( f_{\text{GR}}^{\boldsymbol{\delta}} \left ( \mathbf{H} \right )\right ) $.
The estimator $g_{\text{GR}}^{\boldsymbol{\zeta}}$ takes the output of the feature extractor $f_{\text{GR}}^{\boldsymbol{\delta}} \left (  \mathbf{H} \right )$ as input, and it produces an estimated location $\widehat{\mathbf{P}}$ of the \acp{UE} as output, i.e., $\widehat{\mathbf{P}} = g_{\text{GR}}^{\boldsymbol{\zeta}} \left (  f_{\text{GR}}^{\boldsymbol{\delta}} \left (  \mathbf{H} \right ) \right ) $. 
Now, in \eqref{eq.L_g_GR}, the first term represents localization error similar to the error in \eqref{eq.problem_statement}, and the second term acts as a regularizer. 
That is, $g_{\text{GR}}^{\boldsymbol{\zeta}} \circ f_{\text{GR}}^{\boldsymbol{\delta}}$ in \eqref{eq.L_g_GR} provides location estimates similar to $g$ in \eqref{eq.problem_statement}. 
The relation between \eqref{eq.L_g_GR} and \eqref{eq.problem_statement} is that minimizing the empirical mean in \eqref{eq.L_g_GR} approximates $g$ in \eqref{eq.problem_statement}, and the regularizer in \eqref{eq.L_g_GR} enforces the the extraction of indistinguishable features from wireless fingerprints from both domains in the provided architecture.

To train the components of the proposed architecture above, we consider minimizing the loss function provided in \eqref{eq.L_g_GR}.
Specifically, denoting the minimizer of the function in \eqref{eq.L_g_GR} by $\left ( \boldsymbol{\delta}^{\star}, \boldsymbol{\epsilon}^{\star}, \boldsymbol{\zeta}^{\star} \right )$ and assuming its existence, we attempt to obtain it as follows.
The parameters $\boldsymbol{\delta}^{\star}$ together with $\boldsymbol{\zeta}^{\star}$ are learned using wireless fingerprint - locations pairs from $D_S$, i.e., $\left ( \forall i \in  \left\{1,...,N_S\right\} \right ) \left ( \mathbf{H}_i, \mathbf{P}_i \right ) \in D_S$, by minimizing the $l_2$ distance between true \ac{UE} locations in the source domain $\mathbf{P}_i$ and their estimates $\widehat{\mathbf{P}}_i = g_{\text{GR}}^{\boldsymbol{\zeta}} \left (  f_{\text{GR}}^{\boldsymbol{\delta}} \left (  \mathbf{H}_i \right ) \right ) $.
In contrast, the parameters $\boldsymbol{\delta}^{\star}$ together with $\boldsymbol{\epsilon}^{\star}$ are learned using wireless fingerprint samples from both $\bar{D}_S$ and $\bar{D}_T$, i.e., $\left ( \forall i \in  \left\{1,...,N_S + N_T \right\} \right ) \mathbf{H}_i \in \bar{D}_S \cup \bar{D}_T$, by minimizing the binary cross entropy loss between true domain labels $d_i$ and their estimates $\widehat{d}_i = h_{\text{GR}}^{\boldsymbol{\epsilon}} \left ( f_{\text{GR}}^{\boldsymbol{\delta}} \left ( \mathbf{H}_i \right )\right ) $.
Note that the binary cross entropy loss is scaled by $-\lambda$.
\footnote{The binary cross-entropy loss $\mathcal{L}_{\text{bin}}$ for a single sample is defined as $\mathcal{L}_{\text{bin}} ( d_i, \widehat{d}_i ): = - \left ( d_i \cdot \log ( \widehat{d}_i ) + ( 1 - d_i ) \cdot \log ( 1 - \widehat{d}_i )\right )$, and when $\mathcal{L}_{\text{bin}}$ is multiplied by $- \lambda$, it results in a positive sign for the second term in \eqref{eq.L_g_GR}.}
Such a design of cost function allows $f_{\text{GR}}^{\boldsymbol{\delta}}$ to extract those features from wireless fingerprints from both the source and target domains that improve the performance of the estimator $g_{\text{GR}}^{\boldsymbol{\zeta}}$ while simultaneously degrading the performance of the domain classifier $h_{\text{GR}}^{\boldsymbol{\epsilon}}$, and the parameter $\lambda$ controls the trade-off between the performance of $g_{\text{GR}}^{\boldsymbol{\zeta}}$ and $h_{\text{GR}}^{\boldsymbol{\epsilon}}$.
That is, the extracted features enable the localization of \acp{UE} in the source domain, but it is difficult to distinguish whether these features belong to wireless fingerprints in the source domain or in the target domain.
Therefore, ideally, the estimator trained for localizing \acp{UE} from their wireless fingerprints in the source domain can also be applied in the target domain.

Attempting to obtain parameters $\boldsymbol{\delta}^{\star}, \boldsymbol{\epsilon}^{\star}, \boldsymbol{\zeta}^{\star}$ using conventional optimization methods (e.g., \ac{SGD}) is difficult due to the unconventional update rules for parameters $\boldsymbol{\delta}$, $\boldsymbol{\epsilon}$, $\boldsymbol{\zeta}$ as stated in \cite{Ganin_2015}.
Alternatively, the authors in \cite{Ganin_2015} introduce a \ac{GRL}, enabling us to utilize \ac{SGD} to address challenges associated with unconventional update rules for parameters $\boldsymbol{\delta}, \boldsymbol{\epsilon}, \boldsymbol{\zeta}$.
The \ac{GRL} is inserted between $f_{\text{GR}}^{\boldsymbol{\delta}}$ and $h_{\text{GR}}^{\boldsymbol{\epsilon}}$, as shown in the Fig.~\ref{fig.proposed_methods_diagram} (left), and it has no trainable parameters.
The \ac{GRL} operates differently in forward- and back-propagation, and its behavior can be characterized as follows
\begin{equation*}
	\text{GRL} \left ( \mathbf{x} \right ) = \left\{\begin{array}{cll}
	\mathbf{x} &, \text{if} \hspace{4pt} \mathbf{x} = f_{\text{GR}}^{\boldsymbol{\delta}} \left ( \mathbf{H} \right ) & \text{(i.e., forward-prop.)},\\
	-\lambda \mathbf{x} &,   \text{if} \hspace{4pt} \mathbf{x} = \nabla h_{\text{GR}}^{\boldsymbol{\epsilon}} & \text{(i.e., back-prop.)}.\\
	\end{array}\right.
\end{equation*}
That is, during the forward-propagation, \ac{GRL} acts as an identity operator, and during the back-propagation it takes the gradient from the subsequent level, reverts it by multiplying the gradients with some constant $-\lambda$ and passes it to the preceding level.
With the introduction of the \ac{GRL}, we note $\left ( \forall i \in  \left\{1,...,N_S + N_T \right\} \right )$ $\widehat{d}_i = h_{\text{GR}}^{\boldsymbol{\epsilon}} \Big (\text{GRL}_{\lambda}\big ( f_{\text{GR}}^{\boldsymbol{\delta}} \left ( \mathbf{H}_i \right )\big ) \Big )$ in \eqref{eq.L_g_GR}, and we obtain the parameters $\boldsymbol{\delta}^{\star}$, $\boldsymbol{\epsilon}^{\star}$, $\boldsymbol{\zeta}^{\star}$ by trying to minimize \eqref{eq.L_g_GR} with the \ac{SGD} method.
After obtaining the parameters $\boldsymbol{\delta}^{\star}$, $\boldsymbol{\epsilon}^{\star}$, $\boldsymbol{\zeta}^{\star}$, we discard the domain classifier $h_{\text{GR}}^{\boldsymbol{\epsilon}^{\star}}$, and we use $f_{\text{GR}}^{\boldsymbol{\delta}^{\star}}$ together with $g_{\text{GR}}^{\boldsymbol{\zeta}^{\star}}$ to predict the \ac{UE} locations in the target domain from their wireless fingerprints.
That is, given the problem in \eqref{eq.problem_statement}, we obtain 
\begin{equation*}
	g \approx g_{\text{GR}}^{\boldsymbol{\zeta}^{\star}} \circ  f_{\text{GR}}^{\boldsymbol{\delta}^{\star}}.
\end{equation*}

\subsection{Attention Mechanism in Feature Extraction}
In this work, we implement $f_{\text{AE}}^{\boldsymbol{\alpha}}$ and $f_{\text{GR}}^{\boldsymbol{\delta}}$ with the \ac{SA} mechanism proposed in \cite{Bello_2019, Vaswani_2017}.
The main objective in integrating the self-attention (\ac{SA}) mechanism is to capture the most significant and similar features within wireless fingerprints from both the source and target domains.
This is achieved by providing a weighted sum of the values computed from the input.
Formally, given input $\mathbf{X}$, the concrete steps to implement the \ac{SA} mechanism are as follows:
\begin{itemize}[
		\setlength{\IEEElabelindent}{\dimexpr-\labelwidth-\labelsep}
		\setlength{\itemindent}{\dimexpr\labelwidth+\labelsep}
		\setlength{\listparindent}{\parindent}
	]
	\item Linear transformation of $\mathbf{X}$ into query $\mathbf{Q}$, key $\mathbf{K}$, and value $\mathbf{V}$:
	      \vspace{-4pt}
	      \begin{equation*}
	      	\begin{matrix}
	      		\mathbf{Q} = \mathbf{X} \mathbf{W}_q, & \mathbf{K} = \mathbf{X} \mathbf{W}_k, & \mathbf{V} = \mathbf{X} \mathbf{W}_v, \\
	      	\end{matrix}
	      	\vspace{-3pt}
	      \end{equation*}
	      where $\mathbf{W}_q$, $\mathbf{W}_k$, and $\mathbf{W}_v$ are learnable parameters.
	\item Computation of attention score $\mathbf{A}$ from $\mathbf{Q}$ and $\mathbf{K}$:
	      \vspace{-2pt}
	      \begin{equation*}
	      	\mathbf{A} = \sigma \left( \frac{\mathbf{Q} \mathbf{K}^{\top}}{\sqrt{c}} \right),
	      	\vspace{-2pt}
	      \end{equation*}
	      where $\sigma(\cdot)$ represents the softmax function and $c$ is a normalization factor.
	\item Computation of the weighted sum of the value $\mathbf{V}$ using $\mathbf{A}$:
	      \vspace{-8pt}
	      \begin{equation*}
	      	\mathbf{Z} = \mathbf{A} \mathbf{V}.
	      	\vspace{-5pt}
	      \end{equation*}
\end{itemize}
Following the steps described above, the \ac{SA} mechanism focuses on features within the input data that are more significant to the specified objective.
Note that the \ac{SA} mechanism described above is commonly known as single-head SA. 
In this work, we utilize multi-head \ac{SA}, where each head independently performs its own \ac{SA} simultaneously using distinct linear projections $ \left\{ \left ( \mathbf{W}_q^i, \mathbf{W}_k^i, \mathbf{W}_v^i\right ) \right\}_{i=1}^O$, with $O$ representing the number of heads.
\section{Numerical Results}\label{sec.numerical_results}
\subsection{Simulation Setup}
\begin{figure}[!t]
	\centering
	\centerline{\includegraphics[width=85.0mm]{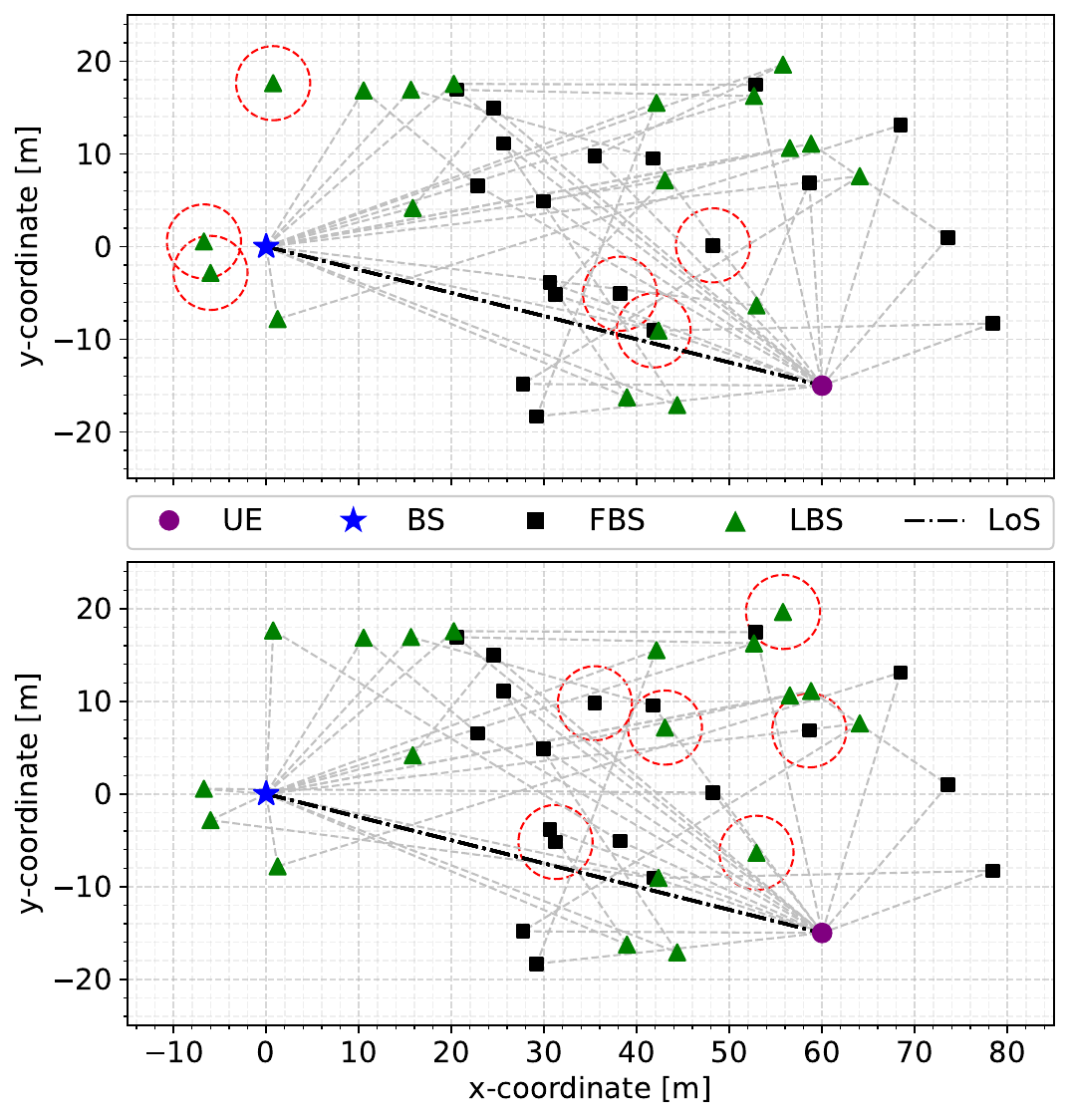}}
	\vspace{-10pt}
	\caption{Simulation of the dynamic indoor \ac{NLoS} environment at $t_1$ (top) and $t_2$ (bottom), each comprising 17 scattering clusters. Each cluster includes a first-bounce scaterrer (FBS) and a last-bounce scaterrer (LBS) through which the transmitted signal from the \ac{UE} reaches the \ac{BS}. The red dashed circle highlights the FBS-LBS pair removed from the total set of 20 clusters.}
	\label{fig.diff_env_cluster}
	\vspace{-15pt}
\end{figure}

{\renewcommand{\arraystretch}{1.3}
	\begin{table}[!t]
		\vspace{7pt}
		\centering
		\resizebox{1.00\hsize}{!}{$
			\begin{tabular}{|c|c|c|}
				\hline
				\textbf{Parameters}                 & \textbf{Value}                                 \\ \hline
				BS antenna dimension {[}\ac{ULA}{]} & $L = 16$                                       \\ \hline
				Antenna spacing {[}wavelength{]}    & $0.5$                                          \\ \hline
				Antenna polarization                & $\pm 45^{\circ}$ (cross-polarization)          \\ \hline
				Localization area {[}m{]}           & $40 \times 40$ centered at $[0,60]$            \\ \hline
				Carrier frequency {[}GHz{]}         & $3.6$                                          \\ \hline
				Bandwidth {[}MHz{]}                 & $20$                                           \\ \hline
				Number of OFDM subcarriers          & $K = 32$                                       \\ \hline
				Propagation scenario                & 3GPP\_38.901\_Indoor\_NLOS \cite{3GPP_2018_06} \\ \hline
			\end{tabular}
		$}
		\vspace{5pt}
		\caption{Simulation parameters.}
		\label{tab:1}
		\vspace{-20pt}
	\end{table}}
To evaluate the performance of the proposed methods in a dynamic wireless environment, we use \ac{QuaDRiGa} \cite{Jaeckel_2016}. 
Specifically, we consider an uplink \ac{NLoS} indoor wireless communication scenario with a single \ac{BS} located at $[0\text{m}, 0\text{m}]$, and multiple \acp{UE} distributed uniformly at random within a $40\text{m} \times 40\text{m}$ square area in a 2D Cartesian coordinate system.
The remaining simulation parameters are provided in Table~\ref{tab:1}.
To simulate realistic wireless channel measurements in dynamic environments, we leverage the spatial consistency feature of \ac{QuaDRiGa} \cite{Kurras_2019}. 
This feature ensures that wireless channel parameters, such as path delays, angles, and power, exhibit similarity for closely located \acp{UE}, with the degree of similarity decreasing as the distance between \acp{UE} increases.
In essence, this feature generates scattering clusters in the environment based on the \ac{UE} locations.
For the \texttt{3GPP\_38.901\_Indoor\_NLOS} scenario used in our simulations, the default number of scattering clusters is 20. 
After generating these clusters, we randomly remove three clusters from the original set, and we repeat this process two more times.
Consequently, we obtain environments at three different time points (ensuring that the same triplet is not removed more than once), denoted by $t_1$, $t_2$, and $t_3$, each comprising 17 clusters.
Furthermore, any pair of environments shares at least 14 clusters in common, as shown in Fig~\ref{fig.diff_env_cluster}.
Thus, $t_1$ represents the time point before the environment change (source domain), while $t_2$ and $t_3$ represent time points after the environment change (target domains), having some similarity with $t_1$ in the sense of Definition~\ref{def.env_similarity}.
\subsection{Simulation}
We evaluate the performance of the proposed methods by comparing them with a baseline method described below.
The baseline method consists of a feature extractor and an estimator. 
The feature extractor is designed to extract features from wireless fingerprints, and the estimator is designed to predict \ac{UE} locations from the extracted features.
In all proposed methods, including the baseline, both the feature extractor and estimator share an identical architecture. 
Specifically, the feature extractor is based on a convolutional neural network, comprising three convolutional layers each with $6 \times 6 $ kernels and $32$ filters, followed by one fully connected layer with $128$ neurons, i.e., $B = 128$. 
The location estimator is based on deep neural network, composed of three fully connected layers with $128$, $64$, and $2$ neurons, respectively. 
The decoder introduced in Sec.~\ref{sec.DA_autoenc} has a symmetrical architecture with a feature extractor.
The classifier introduced in Sec.~\ref{sec.DA_GRL} has an identical architecture with the location estimator with the difference that the classifier outputs values with softmax activation.
The number of \ac{SA} heads is set to $O = 12$.
We train and test all methods with $10 \cdot 10^4$ and $1 \cdot 10^4$ samples, respectively.
The learning rate, number of epochs and batch size are 0.0001, 1000 and 200, respectively.
\begin{figure}[!t]
	\centering
	\centerline{\includegraphics[width=80.0mm]{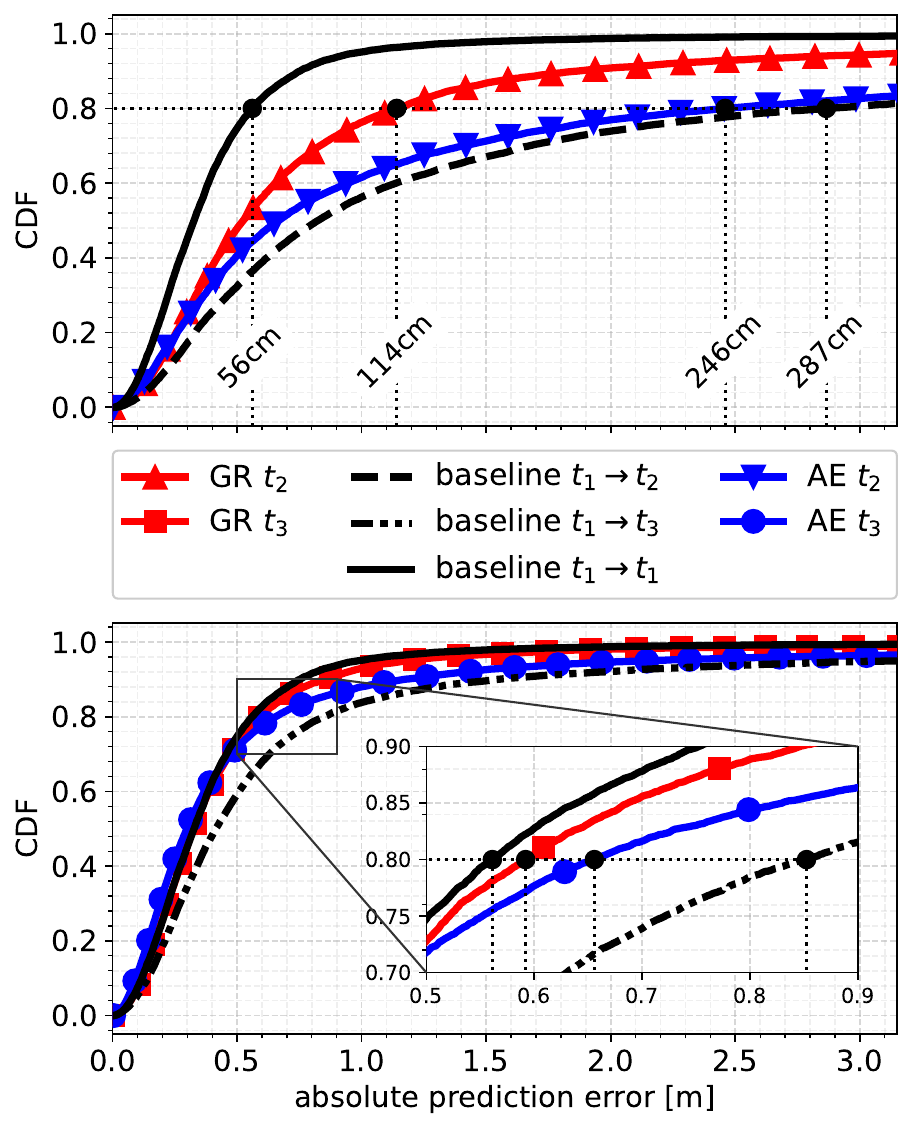}}		
	\vspace{-10pt}
	\caption{Performance of the proposed methods.}
	\label{fig.GR_vs_AE_vs_baseline}
	\vspace{-17pt}
\end{figure}

The performance of the proposed methods is provided in Fig.~\ref{fig.GR_vs_AE_vs_baseline}.
In the provided figures, the x-axis represents the absolute prediction error $ || \mathbf{P} - \widehat{\mathbf{P}} ||$, and the y-axis displays \ac{CDF} or percentiles.
We provide the $80^{\text{th}}$ percentile of the prediction error in Fig.~\ref{fig.GR_vs_AE_vs_baseline}.
The performance of the proposed methods based on the autoencoder and the gradient reversal is denoted as AE and GR, respectively.
In Fig.~\ref{fig.GR_vs_AE_vs_baseline}, AE $t_i$ and GR $t_i$ indicate the performance of the respective methods at $t_i$, and baseline $t_i \rightarrow t_j$ indicates the baseline method trained with samples obtained at $t_i$ and tested at $t_j$.
The numerical approximation of the similarity between environments, in the sense of Definition~\ref{def.env_similarity}, computed with 5000 samples is
\vspace{-3pt}
\begin{equation*}
	\sigma_g \left ( t_1, t_2 \right ) \approx 36.2312, \hspace{10pt} \sigma_g \left ( t_1, t_3 \right ) \approx 31.1494,
	\vspace{-3pt}
\end{equation*}
where $g$ is the baseline estimator trained with fingerprint-location pairs collected at $t_1$.
The approximation above indicates a stronger similarity between $t_1$ and $t_3$ compared to the similarity between $t_1$ and $t_2$. 
This discrepancy is further observed in the performance degradation of the baseline method, as provided in Fig.~\ref{fig.GR_vs_AE_vs_baseline}.

The numerical results demonstrate the capability of the proposed methods to estimate the \ac{UE} locations from their wireless fingerprints after a change in the environment. 
Specifically, the transition from $t_1$ to $t_2$ significantly reduces the performance of the baseline method, while the proposed methods noticeably compensate for the loss of localization performance at $t_2$, as shown at the top of Fig.~\ref{fig.GR_vs_AE_vs_baseline}.
Moreover, the results presented at the bottom of Fig.~\ref{fig.GR_vs_AE_vs_baseline} provide additional insights, indicating a slight performance degradation of the baseline method when transitioning from $t_1$ to $t_3$.
These findings collectively highlight the robustness of the proposed methods against environmental changes, as they consistently outperform the baseline method even in scenarios where there is a marginal decrease in the performance of the baseline estimator.
Thus, the results underscore the practical applicability and effectiveness of the proposed methods in dynamic wireless environments.
\section{Conclusion}
In this work, we proposed methods for localizing \acp{UE} from their wireless fingerprints in dynamic indoor \ac{NLoS} environments. 
The proposed methods consider the environments before and after change as two different domains and utilize an unsupervised domain adaptation approach to obtain an estimator capable of predicting \ac{UE} locations in the environment after the change.
Simulations showed that the proposed methods consistently outperform a baseline estimator trained in the environment before changes and applied in that same environment after the changes occurred.
The results of this work indicate the promising potential of applying the proposed methods for localization in practical wireless \ac{MIMO}-\ac{OFDM} communication systems within dynamic indoor \ac{NLoS} environments.

\end{document}